\documentclass[11pt,preprint,flushrt]{aastex}
\usepackage{amsmath}
\usepackage{times}
\allowdisplaybreaks[1]

\newcommand\etal{{\it et~al.\/}}
\newcommand\cf{{\it c.f.\/}}

\begin{document}

\title{PSF Anisotropy and Systematic Errors in Weak Lensing Surveys}

\author{Bhuvnesh Jain, Mike Jarvis, Gary Bernstein}
\affil{Dept. of Physics and Astronomy, University of Pennsylvania,
Philadelphia, PA 19104}
\email{bjain,mjarvis,garyb@physics.upenn.edu}

\begin{abstract}

Given the basic parameters of a cosmic shear weak lensing survey, 
how well can systematic errors due to anisotropy in the point spread 
function (PSF) be corrected? 
The largest source of error in this correction to date has been the 
interpolation of the PSF to the locations of the galaxies.
To address this error, we separate the PSF patterns into components that
recur in multiple exposures/pointings and those that vary randomly
between different exposures (such as those due to the atmosphere). In
an earlier study we developed a principal component approach to
correct the recurring PSF patterns \citep{Ja05}.
In this paper we show how randomly varying PSF patterns can also be
circumvented in the measurement of shear correlations. For the 
two-point correlation function this is done by simply using pairs 
of galaxy shapes measured in different exposures. 
Combining the two techniques
allows us to tackle generic combinations of PSF anisotropy patterns.
The second goal of this paper is to give a formalism for quantifying
residual systematic errors due to PSF patterns. 
We show how the main PSF corrections improve with
increasing survey area (and thus can stay below the reduced statistical
errors), and we identify the residual errors which do not scale with survey
area. Our formalism can be applied both to planned lensing surveys to
optimize survey strategy and to actual lensing data to quantify 
residual errors.

\end{abstract}

\keywords{cosmology:gravitational lensing --- methods:data analysis}

\section{Introduction}

Weak gravitational lensing refers to the coherent distortions of
background galaxy images by mass structures along the line of
sight. Lensing measurements from imaging surveys have emerged as a
powerful probe of cosmology. With planned surveys that will cover
thousands of square degrees, the statistical errors on measured shear
correlations will be extremely small (e.g. the Dark Energy Survey \citep{DES}, 
PanSTARRS \citep{PanSTARRS}, LSST \citep{LSST},
and SNAP \citep{SNAP}). However systematic errors may
exceed the statistical errors and dominate the error budget on
cosmological parameters.

\begin{figure}[t]
\epsscale{0.6}
\plotone{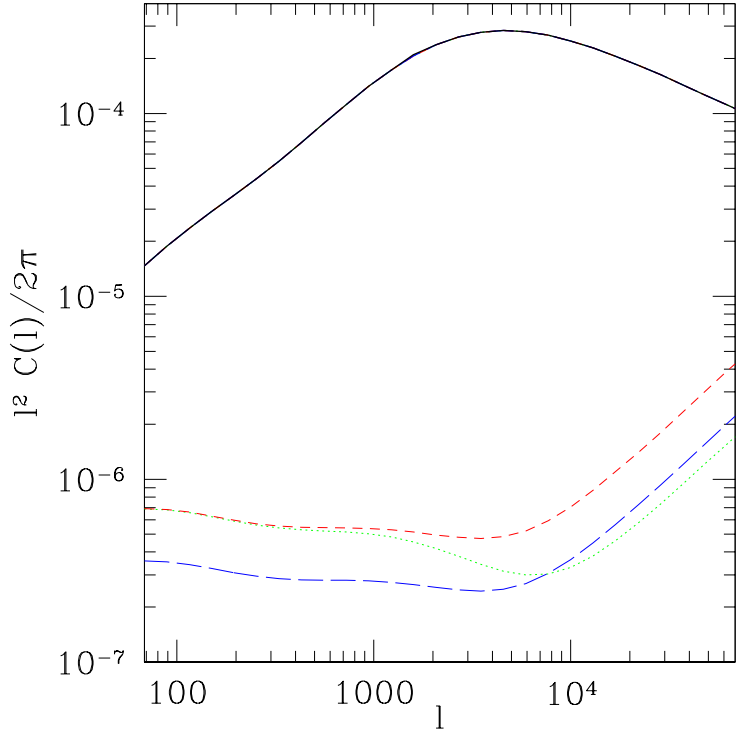}
\caption[]{\small Statistical errors in the lensing power spectrum
for three survey parameters. The upper black curve shows the shear
power spectrum for source galaxies at $z=1$. The three lower curves
show the statistical errors (the sum of sample variance and intrinsic
ellipticity contribution) for three survey parameters described in the
text. Since the PSF anisotropy can be of order
1\%, its contribution ($\sim 10^{-4}$) 
would greatly exceed the statistical errors. 
It must be corrected by several orders of
magnitude to fall below the statistical errors for these survey
parameters.  }
\label{fig:stats}
\end{figure}

To analyze the effect of errors on lensing statistics, we 
will consider the two-point correlation functions of the shear
$\xi_{\gamma+}(\theta), \xi_{\gamma-}(\theta)$ and 
the shear power spectrum $C_\gamma(\ell)$. Other
statistics often used in lensing measurements, such as the aperture
mass variance and the top-hat shear variance, can be obtained by
integrating the two-point correlation functions \citep{Sch02}, so
there is no need to consider them in addition.

The contribution to the shear correlations per log interval in $\ell$ is
$\ell^2 C_\gamma(\ell)/2\pi$, which makes it an intuitive way 
to plot the shear power spectrum.  Figure \ref{fig:stats} shows this power
spectrum for the (nonlinearly evolved) concordance $\Lambda-$CDM
model, and the statistical errors on it for three choices of survey
parameters. The statistical errors include sample variance, which
dominates on large scales ($\ell$ below a few thousand) and the shot noise
contribution of the intrinsic ellipticites of galaxies. 
The systematic errors should be smaller than the sum of these 
so as not to dominate the error budget.  

The upper black curve shows the
shear power spectrum for source galaxies at redshift $z=1$.  The
lowest statistical errors (long-dashed blue curve) are for a survey
similar to that planned for the LSST, which covers half the
sky: $f_{\rm sky}=0.5$, $n_g=40$ with rms intrinsic contribution to
the shear $\sigma_\gamma=0.25$. The dashed red curve is for a ground
based survey with smaller sky coverage: $f_{\rm sky}=0.1$, $n_g=40$.
The dotted green curve shows the statistical error for a space based
survey with $f_{\rm sky}=0.1$, $n_g=100$. The higher number density
reduces the error at high $\ell$ compared to the ground based survey
with the same sky coverage.  For $\ell > 10^4$, on sub-arcminute
scales, the statistical error rises due to the intrinsic ellipticity 
contribution which has a white noise power spectrum (assuming the
intrinsic ellipticities are uncorrelated and randomly oriented). However the
statistical errors are roughly constant over the range of scales that
provide the cosmological information ($100 < \ell < 10^4$). This is
useful for setting the permissible level of residual systematics. 

One of the main sources of error in the shear estimates comes from the
convolution of the image by the point spread function (PSF).  
This function is known (albeit noisily) at the positions of the stars 
in the image. As the PSF varies across the image, one must interpolate 
this function to the positions of the galaxies.
An incorrect model of the PSF leads to an
error in the estimated (pre-seeing) galaxy shape and hence in the shear
correlation. Coherent PSF patterns have non-zero two-point functions,
which add to the lensing induced correlations in galaxy
ellipticities. This systematic error can exceed statistical errors in
lensing measurements if the PSF is not modeled sufficiently
accurately \citep{Ho04}.
We will not consider here errors due to removal of the PSF 
from the galaxy shapes if the PSF at the location
of the galaxy is known correctly \citep[e.g.][]{KSB,Ka00,BJ02,RB03}.  
We are only concerned here with the estimation of the PSF at each galaxy's
location.

PSF interpolation error has been one
of the primary sources of systematic error in most of the 
lensing measurements published to date
(errors in the shear calibration and redshift distribution are the 
other main sources). 
Given a model for PSF
anisotropy we can calculate how well the power spectrum would need to
be corrected to be well below the statistical errors.  
The statistical error curves in 
Figure \ref{fig:stats} give a good indication of the upper limit
on coherent residual systematic errors if they are not to dominate the
error budget. Thus at $l\sim 1000$, or 10 arcminute scales, the
coherent residual should be well below $0.001$ (so that its square is
smaller than the statistical error curves).  Generic models of PSF
patterns do not exist; the amplitudes measured in
current data (before any corrections) 
are in the range 1-10\% with varying coherence
scales. Telescopes that will be built with lensing as a primary science goal
are expected to do better than these, and may have PSF modeling
software like TinyTim for HST, but even for the best-designed
telescope, the galaxy shapes will require correction using data on stars.

In this paper we describe two methods which in combination can remove
the systematic effects of asymmetric PSFs in large imaging
surveys. A method based on a principal component analysis (PCA)
of the PSF was the subject of a
recent paper \citep{Ja05}.  Essentially, it detects and models 
components of the PSF pattern which appear in many different images.
For example, guiding errors have the same effect on every star in 
an exposure, so its pattern is a constant in $(x,y)$, with a 
coefficient which varies from exposure to exposure.  The principal
component corresponding to this is therefore a constant.
Focus errors are similarly recurring; astigmatism produces a characteristic
pattern when the telescope is slightly above focus, and the opposite
pattern when below.  That is, there is a fixed $(x,y)$ pattern which
is modulated by a coefficient for each exposure.
(There may be more than one principal component 
corresponding to focus if the variation is not quite linear as the 
telescope gets more out of focus.)  
In general, the principal components
should model any pattern due to a recurrent physical cause.
The second method discussed in this paper
tackles PSF patters that no not recur in different exposures. 

In \S\ref{pca}, we outline the pipeline for lensing measurements to
show where different sources of error enter. We quantify the 
residual systematic errors due to
the PSF pattern after performing the PCA interpolation and show how
they scale with
survey area.  We find that some components of the error scale as ${\rm
Area}^{-1}$.  However, other
components of the error remain roughly constant even with this
interpolation scheme. We discuss how these components are ones 
that do not recur across different exposures. 

In \S\ref{multiimage}, we show how to completely eliminate these
non-recurring systematic errors by correlating only
galaxy shapes measured on distinct exposures.  Since
the two- and three-point correlation functions encompass most of the
lensing information that will be desired from current and future
surveys, the combination of these two techniques will lead to the near
elimination of systematic errors stemming from PSF interpolation.

In \S\ref{discussion} we discuss the key requirements for a survey to keep
residual systematics sufficiently small. We list the ingredients that
determine these residuals and describe how to estimate them for
planned surveys as well as from actual survey data. Future work needed
to test and refine this approach is discussed. 

\section{Systematic Errors Due to PSF Interpolation}
\label{pca}

\subsection{The Lensing Pipeline}

We begin by outlining the pipeline used to estimate lensing 
statistics from images of the sky. This will help us identify the
steps at which different errors enter and how new techniques
can reduce certain errors. We will introduce the PCA technique below in 
step \ref{interpstep} on PSF interpolation 
and use it in subsequent sections to quantify residuals. 

The lensing pipeline can be summarized in 
\ref{npipesteps} main steps:

\begin{enumerate}

\item{\em Detection of Stars and Galaxies}

Weak lensing surveys generally observe the same portion of the sky on
several separate exposures.  Each of these exposures are usually made
up of multiple images, from the multiple CCD chips in the camera.
These factors can make object detection and measurement somewhat
complicated.  Generally, one wants to stack all of the images for a
given part of the sky to get the best signal-to-noise for detecting
objects.  
However, if the shapes are measured from the stacked image,
there are issues due to correlated noise from the image-combining algorithm,
and even slight registration errors can lead to very significant errors in
the shape measurements.  Also, the PSF on the stacked image will be
near the middle of the range of seeing values, so the signal-to-noise
for the smallest galaxies may actually be worse on the stacked image
than on the best-seeing images.  Worse, the PSF pattern on the stacked
image will change abruptly at the edge of every input exposure, so if
there are large offsets in the original pointings, the PSF pattern
will be impossible to model precisely.

Therefore, we generally recommend detecting objects on a stacked
image, but measuring the PSF and galaxy shapes on the images from the
individual exposures.  This will allow for good PSF interpolation, and
the keep the pixel noise uncorrelated.  If the image registration is
good enough, one can centroid on the stacked image and use it
for the individual measurements, which may improve the signal-to-noise
of the shape estimates.

In \S\ref{multiimage} we point out that using shapes from different 
exposures eliminates systematic errors due to certain PSF patterns; 
this also argues against stacking images to measure shapes. It will 
impact the optimal number of images to observe per location, which 
we discuss \S\ref{exposures}. 

For surveys with very many (more than about 10 or 20) exposures at
each location, it may make sense to stack together several exposures
which have roughly the same seeing and which are not (very) offset
from each other.  Each location would then have a smaller number of
images for measuring shapes.  For the purposes of this paper, the term
exposure would then refer to these stacked images, rather than the
original exposures.

Finally, we remind the reader that the PSF can act as a matched filter for
galaxies that are aligned in the same direction as it. This selection 
bias can introduce a systematic error.  The error is eliminated by 
detecting galaxies which
are as faint as possible, and then selecting according to an 
shape-independent signal-to-noise estimate.  
A similar error, which is more difficult 
to remove, is that galaxies are more likely to be blended along the 
direction of the PSF, which will bias galaxy shapes in the same direction.

\item{\em Measurements of the PSF}

After identifying stars and galaxies in an image, the stars are
used to measure various aspects of the PSF.  Different analysis
methods use different components of the PSF, but all methods
measure the ellipticity and size at least.  More sophisticated
analyses require some higher order shape information as well.  Since
the PSF generally varies between different exposures as well as 
across the field, these values are measured as a function of the 
positions $\vec \phi^{(i)}$ of stars in exposure $i$. 

There are three systematic errors which may be introduced at this stage.
First, small galaxies may be falsely identified as stars, which will
lead to errors in the PSF estimates.  Second, if the PSF is color
dependent, the PSF measured by the stars may not be (exactly) the same
as the PSF which has acted on the galaxies.  This color error may
be redshift dependent which would complicate tomography analyses.
Similarly, if the detector response is slightly non-linear, the PSF
of the bright stars may be different from the PSF of the faint galaxies.

\item{\em Interpolation of the PSF}
\label{interpstep}

We need to know the PSF at the location of the galaxies, which are 
the tracers of the lensing shear.  Since the PSF is not measured at these
locations, we need to interpolate the measurements
from the locations of the stars.  Here, we briefly
describe how to do this using the principal component analysis (PCA)
method (described in greater detail in \citealp{Ja05}):

For each exposure, $i$, we find a polynomial fit to the PSF measurements,
$Q_i^{(m)}(\vec\phi^{(i)})$, 
where the order $m$ is given by the number of stars
available for fitting: $(m+1)(m+2)/2 \leq N_*$,
and $\vec\phi^{(i)}$ is the position measured in the coordinate system
of exposure $i$. 
In practice, one may want
to use a separate polynomial for each chip to avoid smoothing over
discontinuities at the chip boundaries.

Next we take the patterns for all of the exposures and pointings, 
as quantified by the coefficients in the polynomials $Q_i$, and find 
the principal
components of the variation.  This will find the patterns that repeat
over a significant number of the exposures.  We sort the principal
components according to how much they contribute to the total
variation of the PSF patterns (i.e. the singular values).
At some point the components will not
be important for describing the patterns, so we choose some cutoff and
only use the first $N_{\rm PC}$ components.  (We will discuss how to
determine this number below.)  
Thus, the PSF pattern for each exposure is described as a weighted sum
of the various principal components:
\begin{equation}
e_{\rm PSF}(\vec \phi^{(i)}) = 
  \sum_{k=1}^{N_{\rm PC}} a_{ik} P_k(\vec \phi^{(i)})
\label{eqn:psf}
\end{equation}
where $i$ is the index number of the exposure, $e$ can represent the
PSF ellipticity, or any other feature of the PSF, such as size or any
higher order shape information, and $P_k$ is the $k$th principal component
for the same quantity.

We can then refine these principal components, $P_k$, using a higher order
polynomial, by keeping the $a_{ik}$ coefficients fixed and
using the stars in all of the exposures for the fit.
For very large surveys, this allows us to use significantly 
higher order polynomials which more accurately describe the components.

\item{\em Measurements of the galaxy shapes}

Given the relevant description of the PSF at the location of a galaxy,
one can make an estimate of the galaxy's shape before convolution by
the PSF.  Our methods for doing this are described in \citet{BJ02},
but there are other methods for this step as well.
For the purposes of this paper, we will assume that the only errors
introduced here are the measurement noise and the intrinsic shape
noise of the galaxy.  That is, if the knowledge of the PSF were
perfect, we assume that this step would then produce perfectly
unbiased estimates of the shear at each galaxy's location. 

In reality, there may be systematic errors due to the dilution 
correction (the effect of the size of the PSF on galaxy shapes; 
\cf \ \citealp{Hi03}) or the shear calibration
(the response of the distribution of galaxy ellipticities 
to the shear). These errors lead to multiplicative errors in the shear
two-point correlations. We discuss in \S\ref{discussion} how the 
improved PSF interpolation we describe would reduce the 
dilution errors as well. Recent studies \citep{Hu05,Gu05} show that
the impact of these errors on cosmological parameter estimation is
less severe than the additive errors due to PSF anisotropy, 
as they can be self-calibrated from the data.. 

\item{\em Correlation of the shear estimates and comparison to theory}
\label{npipesteps}

The lensing shear information is contained primarily in the two- and
three-point shear correlation functions.  In fact, most other shear
statistics, such as the shear variance and the aperture mass variance
and skewness, can be expressed as integrals over these functions. 
Exceptions include the convergence probability distribution
function \citep{Zh05}, peak statistics \citep{JvW,Mi02}
and topology measures \citep{Ma01,Sato01,Sato03},
which contain more information about the shear than is contained
in its low-order correlations. 

There are two two-point shear correlation functions:
\begin{align}
\xi_+(\theta) &= \left\langle \gamma(\vec\phi)
\gamma^*(\vec\phi+\vec\theta) \right\rangle \\ \xi_-(\theta) &=
\left\langle \gamma(\vec\phi) \gamma(\vec\phi+\vec\theta)
\right\rangle
\label{eqn:two-point}
\end{align}
where $^*$ indicates complex conjugate of the complex-valued shear
estimate, and the shears are measured relative to the line joining the
two galaxies.  Likewise, there are four three-point shear correlation
functions which are a function of the size and shape of the triangle
connecting the three galaxies.

Since other statistics may be derived from these, we take the
correlation functions to be the final product of the lensing pipeline
which is then used to constrain cosmology.  For tomography
applications, the correlation functions are measured as functions of 
redshift bins as well as the angular separation.  The following
discussion of the errors would then refer to the errors in the 
correlation functions for each pair (or triplet) of redshift bins.

The estimation of cosmological parameters from the measured shear
correlations relies on the use of redshift information. 
Errors in the estimated redshift bins (or the overall distribution 
of redshifts for non-tomographic applications) are an
important systematic error which may be introduced at this point.
Accurately calibrating the redshifts is a big concern for upcoming
large cosmic shear surveys \citep{Ma05,Hu05}. Spectroscopic
sub-samples that extend to high redshifts 
may be necessary to calibrate the redshift distribution, 
though \citet{Mandel05} show how to some extent it can be calibrated 
from the data.  

There may also be systematic errors introduced by the theoretical
predictions.  In particular, current estimates 
of the non-linear power spectrum may have errors of order 5\% at 
scales of several arcminutes \citep{Sm03}, or even larger for 
quintessence models \citep{Kl03}. (See \citealp{Linder05} for 
an improved prescription for generic dark energy cosmologies.)
On small scales it is also 
necessary to consider baryonic physics \citep{White04,Zh04}
and higher order effects, 
e.g. to account for the fact that galaxy shape measurements estimate the
reduced shear $g=\gamma/(1-\kappa)$, not the shear $\gamma$ 
directly \citep{White05,Do05}.
Theoretical predictions which do
not correctly take this into account would introduce an error
on small scales. 

\end{enumerate}

We have seen that there are systematic errors which may be introduced
in every step of the lensing pipeline.  The errors from the PSF 
interpolation step
have often been considered the most difficult to remove due to the 
limited number of stars per exposure.  The most problematic of 
the other errors are the shear calibration and the redshift calibration.
There is ongoing work aimed at limiting the impact of these errors
on cosmological parameter estimation from future lensing data. 
However, they are not the focus of this paper; henceforth we restrict
our discussion to the PSF interpolation errors.

\subsection{Residual Systematic Errors after PCA Interpolation}

We now want to determine what the residual systematic errors in the
estimates of the correlation functions are, due to imperfect PSF
interpolation.  Let the estimated shear in exposure $i$ at position
$\vec \phi^{(i)}$ be
\begin{equation}
\hat{\gamma}(\vec \phi^{(i)}) = \gamma_{\rm grav}(\vec \phi^{(i)}) +
\epsilon_{\rm intrinsic}(\vec \phi^{(i)}) + \epsilon_{\rm meas}(\vec
\phi^{(i)}) + \epsilon_{\rm PSF}(\vec \phi^{(i)})
\label{eqn:psf_error}
\end{equation}
where $\gamma_{\rm grav}$ is the lensing signal, $\epsilon_{\rm
intrinsic}$ is the noise due to the intrinsic shape of the
galaxies, $\epsilon_{\rm meas}$ is the statistical error in the
measurement from the photon shot noise, and $\epsilon_{\rm PSF}$ is
the error in the shear estimate due to uncorrected PSF contamination.

The systematic errors from the PSF interpolation enter through
the term $\epsilon_{\rm PSF}$, which arises from errors in the
PSF ellipticity. Using equation \ref{eqn:psf}, which
expands it in principal components, we can write $\epsilon_{\rm PSF}$ as:
\begin{equation}
\epsilon_{\rm PSF}(\vec\phi^{(i)})
= \frac{1}{\cal R} \sum_{k \le N_{\rm PC}} \left [ 
\delta a_{ik} P_k(\vec \phi^{(i)})
+ a_{ik} \delta P_k(\vec \phi^{(i)}) 
\right ]
+ \frac{1}{\cal R} \sum_{k>N_{\rm PC}} a_{ik} P_k(\vec \phi^{(i)}) 
\label{eqn:psf2}
\end{equation}
where $\delta$ refers to the error in the estimate of a quantity, and
the last sum includes all of the patterns which are not modeled by 
the PCA, including any completely random effects which do not
recur in multiple exposures. ${\cal R}$ represents the conversion from
ellipticity to shear, which \citet{BJ02} refer to as responsivity\footnote{
More generally, ${\cal R}$ is the net effect on the shear estimate
due to an error in the PSF ellipticity, which may include other
effects than that described by the ${\cal R}$ of \citet{BJ02}.}.
When using this technique for other properties of the PSF besides 
ellipticity (size for example), ${\cal R}$ would
be the corresponding mean effect that errors in the measurement have 
on the net shear estimates from the galaxy shapes.

We will refer to the estimates of the two-point correlation function
from observations of galaxies on two exposures, $i$ and $j$ as:
\begin{equation}
\hat\xi^{(i,j)}(\theta) = \left\langle \hat\gamma(\vec\phi^{(i)})
  \hat\gamma(\vec\phi^{(j)}+\vec\theta) \right\rangle
\label{eqn:xi_ij}
\end{equation}

where we omit the $+$ and $-$ subscripts, both here and in much of the
further discussion, leaving the appropriate conjugation or not in the
two cases implied.

The statistical errors from the measurement noise and intrinsic
ellipticities are well understood.  We now look at what can contribute
to the systematic PSF contamination, $\epsilon_{\rm PSF}$, and how
that propagates to $\hat\xi^{(i,j)}$.  The errors in the three-point
function are completely analogous, so it is sufficient to only refer
to the two-point function here.

\begin{enumerate}

\item{\em Errors in the principal components, $P_k$}
\label{error_p}

There will be errors in the estimates of the $P_k$ functions due to
the simple fact that we constrain them with a finite number of stars.
These lead to systematic errors in the correlation functions, since we
use the same principal components for all of the exposures, so the
errors in the $P_k$ repeat for every pair of galaxies that is used to
calculate the correlation function.

If the error in each estimated $P_k$ is $\delta P_k$, then the
propagated error in the correlation function due to this is (from
Equations~\ref{eqn:psf2} and \ref{eqn:xi_ij}):
\begin{equation}
  \delta \hat\xi_{(\ref{error_p})}^{(i,j)} = \frac{1}{{\cal R}^2}
  \sum_{k,\ell \le N_{\rm PC}}
    a_{ik} a_{j\ell}
    \left\langle \delta P_k(\vec\phi^{(i)}) 
      \delta P_\ell(\vec\phi^{(j)}+\vec\theta) \right\rangle 
\label{eqn:pc_error}
\end{equation}

\item{\em Unmeasured principal components}
\label{error_unmeasured}

The above analysis only included a finite number of principal
components, $N_{\rm PC}$.  Any PSF variation that is described by
components of lower significance than these has been completely
unmodeled.  Therefore, all of this PSF power will still be uncorrected
in the galaxy shapes, which will lead to a systematic error in the
shear estimates:
\begin{align}
\nonumber
\delta \hat\xi_{(\ref{error_unmeasured})}^{(i,j)} &= \frac{1}{{\cal R}^2} 
  \sum_{k>N_{\rm PC}} \Bigg( 
    2\ \sum_{\ell \le N_{\rm PC}} 
    a_{ik} a_{j\ell} 
    \left\langle P_k(\vec\phi^{(i)}) 
      \delta P_\ell(\vec\phi^{(j)}+\vec\theta) \right\rangle  \\
&\qquad \qquad \quad \quad
  + \sum_{\ell>N_{\rm PC}} 
    a_{ik} a_{j\ell}
    \left\langle P_k(\vec\phi^{(i)})
      P_\ell(\vec\phi^{(j)}+\vec\theta) \right\rangle 
  \Bigg) 
\label{eqn:unmeasured1} \\
&\approx \frac{1}{{\cal R}^2}
  \sum_{k>N_{\rm PC}}
    a_{ik} a_{jk}
    \left\langle P_k(\vec\phi^{(i)})
      P_k(\vec\phi^{(j)}+\vec\theta) \right\rangle 
\label{eqn:unmeasured}
\end{align}
where $a_{ik}$ refers to the values that the coefficients for the
unmeasured components would have if they were included in the
analysis.
The dominant terms in this expression will typically be the 
autocorrelation terms as written in Equation~\ref{eqn:unmeasured}; 
however,
it is possible that there could be significant correlations with
either other unmeasured components or the errors in the measured 
components as shown in Equation~\ref{eqn:unmeasured1}.

\item{\em Non-recurring contributions to the PSF pattern}
\label{error_atm}

Some portion of the PSF pattern is completely random and uncorrelated
between different exposures, e.g.\ atmospheric effects. These
non-recurring contributions will remain as a systematic error in the
shear correlations since they can have spatial structure. \citet{Wi05}
has recently measured the atmospheric contribution to PSF
errors from the Subaru telescope, while \cite{KTL00} modeled its 
spatial and temporal coherence. The actual level of atmospheric contribution
will depend on details of the instrument and observing strategy. Here we
will consider the atmosphere as well as non-recurring contributions from
the instrument in one category of PSF errors. 

These could be viewed as a subset of the unmeasured principal
components described above, since the information here must be
contained in a complete principal component analysis with $N_{\rm PC}$
equal to the total number of exposures.
These uncorrelated contributions would be described
by the myriad very low significance components which constitute most
of those neglected by using a (much) lower $N_{\rm PC}$. 
However, we choose to make
them a separate item to point out that these contributions are
completely uncorrelated from one exposure to another, which make them
a qualitatively different type of error.  In particular, this means
that $\langle a_{ik} a_{jk} \rangle = 0$ for $i \neq j$, so the
systematic error only occurs for estimates of $\xi$ with $i = j$:
\begin{equation}
\delta \hat\xi_{(\ref{error_atm})}^{(i,i)} = \frac{1}{{\cal R}^2} 
  a_{{\rm atm},i}^2 
  \left\langle P_{\rm atm}^{(i)} (\vec\phi^{(i)}) 
    P_{\rm atm}^{(i)}(\vec\phi^{(i)}+\vec\theta) \right\rangle
\end{equation}
where $P_{\rm atm}^{(i)}$ is the portion of the PSF pattern which is uncorrelated
with that from any other exposure.

\item{\em Errors in the coefficients $a_{ik}$}
\label{error_a}

There will be errors in the estimates of the $a_{ik}$ coefficients as
well, since they will be constrained by the finite number of stars in
each exposure.  These errors lead to systematic errors in the
correlation function, since the amount of correction on the galaxies
for each principal component will be slightly wrong.  This will then
add a little bit of the correlation functions of the components to the
shear correlation function:
\begin{equation}
\delta \hat\xi_{(\ref{error_a})}^{(i,i)} = \frac{1}{{\cal R}^2}
  \sum_{k,\ell \le N_{\rm PC}}
  {\rm Cov} \left(a_{ik}, a_{i\ell}\right) 
  \left\langle P_k(\vec\phi^{(i)}) 
    P_\ell(\vec\phi^{(i)}+\vec\theta) \right\rangle
\label{eqn:coeff}
\end{equation}
Note that, like the previous systematic, this systematic is nonzero
only for estimates of $\xi$ with $i = j$, since the errors on the coefficients
are uncorrelated between exposures.

\end{enumerate}

\subsection{Scaling of Systematics with Survey Parameters}
\label{residuals}

Surveys with degree sized fields of view (FOV), covering total area of
1000 square degrees or larger, are likely to have enough exposures and
stars to make accurate corrections to PSF anisotropy. Here we
quantify the residual PSF systematics
described above and discuss their possible impact on survey strategy.

The following parameters constitute our description of a lensing
survey.
\begin{list}{}{}
\item Field of view, in steradians: $\Omega_{\rm FOV}$
\item Number of pointings: $N_{\rm point}$
\item Number of exposures per pointing: $N_{\rm exp}$
\item Mean number of stars per exposure: $\langle N_* \rangle\equiv
N_* = n_* \ \Omega_{\rm FOV}$
\item Number of significant principal components: $N_{\rm PC}$
\end{list}{}{}

The survey size is given by the number of pointings as: 
$\Omega_{\rm S} = N_{\rm point} \Omega_{\rm FOV}$.
For PSF measurement $N_*$ includes only those stars that have well
measured shapes, and which are robustly identified as stars.
Interloping small galaxies are an additional concern if one tries to
push the stellar locus too close to the galaxy locus of the
size-magnitude diagram. 
The number of significant principal components used to describe the
recurring PSF pattern, $N_{\rm PC}$, will not likely be known in advance 
of the data.  In fact, it is still variable after obtaining data; 
we describe how to determine a good value for it in \S\ref{scale12}.

We will make the simplifying and conservative assumption that all
exposures in a given part of the sky are centered on the same point,
so they do not sample the PSF on different parts of the
camera\footnote{ 
If this is not true, and the exposures are offset
from each other, then the total number of PSF measurements increases
by a factor of $N_{\rm exp}$, since the stars in every exposure give a
new sample of the PSF patterns.  If the exposures have the same
pointing, then the extra exposures per location do not provide
additional constraints on the PSF pattern.  }. 
Hence the total number
of PSF measurements is $N_* N_{\rm point}$.  These are used to measure
the $N_{\rm PC}$ principal components.  The maximum order of the
polynomial that can be used for each principal component is then of
roughly $(N_* N_{\rm point}/N_{\rm PC})^{1/2}$, which can be much
larger than the order possible by using just the stars in a single
exposure.

\subsubsection{Scaling of Errors 1 and 2}
\label{scale12}
The magnitude of the errors in the principal components scales
according to the total number of stars used to constrain each
component.  Since all of the stars in the survey need to jointly
constrain $N_{\rm PC}$ components, we have
\begin{equation}
  |\delta P_k(\vec\phi)| \propto 
    \left( \frac{N_{\rm PC}}{N_* N_{\rm point}} \right)^{1/2}
\end{equation} 
The systematic error $\delta \hat\xi_{(\ref{error_p})}^{(i,j)}$ will
then scale as 
\begin{equation}
\delta \hat\xi_{(\ref{error_p})}^{(i,j)} \propto 
  N_{\rm PC} \ |\delta P(\vec\phi)|^2 \propto 
    \frac{N_{\rm PC}^2}{N_* N_{\rm  point}} 
\end{equation}
since each element in the sum for $\delta \hat\xi_{(\ref{error_p})}^{(i,j)}$ 
in Equation~\ref{eqn:pc_error}
is quadratic in the $\delta P_k$ functions (assuming that the
cross-terms involving different principal components are negligible).  
As survey size increases (increasing $N_{\rm point}$), this systematic
error will decrease even faster than the statistical errors in
the shear, which decrease as $N_{\rm point}^{-1/2}$, 
presuming that the PC patterns do not evolve as the survey progresses.

The error due to the neglected principal components, $\delta
\hat\xi_{(\ref{error_unmeasured})}^{(i,j)}$, will only decrease if we
increase $N_{\rm PC}$, since the largest neglected components will
then have smaller rms amplitude.  However, increasing $N_{\rm PC}$
will increase the previous error, since each principal component will
be less well measured.  In an ideal analysis, the number of principal
components would be set so that the systematic errors for each of
these two factors is stationary in the number of components.  That is,
the improvement due to adding an additional component should exactly
offset the loss due to the other components being slightly less well
measured.  In general, it is not easy to determine at what 
$N_{\rm PC}$ this will happen.
One needs to look at some measure of the
contamination as a function of $N_{\rm PC}$ to find the minimum total
contamination.  We discuss a few such measures in \S\ref{discussion}.

If such a procedure is done, then the amplitude of the first
neglected component will scale approximately as 
\begin{equation}
|P_{N_{\rm PC}+1}(\vec\phi)| \propto 
  \left( \frac{N_{\rm PC}}{N_* N_{\rm point}} \right)^{1/2}.
\end{equation}
Furthermore, in our CTIO survey data,
we have found that the large-$k$ asymptotic behavior of $|P_k|$ is
\begin{equation}
|P_k(\vec\phi)| \propto e^{-\alpha k}
\end{equation}
with $\alpha$ of order $0.02$.
Therefore, we can estimate $\delta \hat\xi_{(\ref{error_unmeasured})}^{(i,j)}$ 
as
\begin{align}
\delta \hat\xi_{(\ref{error_unmeasured})}^{(i,j)}
&\propto \sum_{k>N_{\rm PC}} |P_k(\vec\phi)|^2 
  \approx \frac{1}{2 \alpha} |P_{N_{\rm PC}+1}(\vec\phi)|^2 \\
&\propto \frac{N_{\rm PC}}{\alpha N_* N_{\rm point}} 
\end{align}
Assuming the asymptotic behavior is relatively generic and that the
necessary increase in $N_{\rm PC}$ occurs significantly
more slowly than the increase in $N_{\rm point}$, both
systematics would scale as $N_{\rm point}^{-1}$.

However, we should point out that one is also limited by the
constraint that $N_{\rm PC} < N_*$, otherwise the coefficients
$a_{ik}$ cannot be measured: so if too many principal components
become important, it will eventually become impossible to include all
of them, and $\delta \hat\xi_{(\ref{error_unmeasured})}^{(i,j)}$ will
not scale as $N_{\rm point}^{-1}$ any further.  Thus, it is important
in designing a survey to try to minimize the number of sources of 
PSF variation to keep the number of principal components reasonably
low.  

\subsubsection{Scaling of Errors 3 and 4}
The error from the atmosphere's PSF, $\delta
\hat\xi_{(\ref{error_atm})}^{(i,i)}$, is essentially constant with
survey size.  Some of the atmosphere's PSF pattern will be modeled by
the various principal components, but most of the PSF power will
remain as a systematic error, especially the high order power, which
will almost always be completely different from the high order power
of the PC's.  In particular, the atmosphere's pattern on scales
smaller than the stellar separation cannot be modeled by the PCA 
or any other method.
The magnitude of this contribution seems to
be relatively small for current surveys.  But for upcoming larger
surveys, its contribution may become dominant over the
residuals from the coherent patterns due to the telescope.
Space-based surveys will not have this contribution, although it is
possible that they will have
other sources of PSF patterns which are uncorrelated between exposures.

The errors in the coefficients $a_{i,k}$ do not scale with the number
of pointings, since they are constrained only by the stars in a single
exposure.  For each exposure, $i$, there are $N_*$ stars which are
used to constrain $N_{\rm PC}$ coefficients.  Take the stars in an
exposure to be numbered $m = 1..N_*$ with positions $\vec \phi_m$,
shapes $e_m$, and shape uncertainties $\sigma_m$.  Also, define $A$ to
be the vector of coefficients for that exposure ($A_k = a_{i,k}$),
define a vector $E$ with $E_m = e_m/\sigma_m$, and define a matrix $Q$
with $Q_{m,k} = P_k(\vec \phi_m)/\sigma_k$.  Then the least-squares
solution for $A$ is:
\begin{align}
\label{qtq}
A &= (Q^T Q)^{-1} (Q^T E) \\ {\rm Cov}(A) &= (Q^T Q)^{-1}
\end{align}

If $N_* < N_{\rm PC}$, then $Q^T Q$ is singular and the errors on A
are infinite.  If $N_* = N_{\rm PC}$ then there is some combination of
coefficients whose error is proportional to the largest $\sigma_m$.
If we sort the stars by $\sigma_m$, so that the largest $\sigma_m$ is at
$m = N_*$, then
\begin{equation}
\sigma(\tilde a) = K \sigma_{N_*}
\end{equation}
where $\tilde a$ is the least well measured linear combination of
coefficients and $K$ is a constant which depends on the values of
$P_k(\vec \phi_m)$.  $K$ is generally of order the rms value of
$P_k(\vec\phi_m)$, but it can be arbitrarily larger for unfortunate
sampling of the principal components\footnote{ 
For example, if all of
the stars happen to be where some component has very little power,
then they will not be able to constrain the coefficient of this
component very well.  }. 
Finally, if $N_* > N_{\rm PC}$ (the usual case), then it can be shown that
\begin{equation}
\sigma(\tilde a) \ge K' \left ( \sum_{k \ge N_{\rm PC}}
  \frac{1}{\sigma_k^2} \right )^{-1/2}
\end{equation}
where the sum is over the $(N_*-N_{\rm PC}+1)$ least well-measured
stars\footnote{
The proof of this expression is somewhat technical, but we direct the 
interested reader to \citet{Golub}, p. 443.  The derivation of our formula
is based on their proof of Theorem 8.5.3 regarding singular values of a 
diagonal matrix plus a rank-1 matrix.
}.
Since $\sigma_k \propto 1/\nu$, where $\nu$ is the
signal-to-noise of the star, the sum in the above formula is dominated
by the highest signal-to-noise stars.  Thus the overall error scales
roughly as the shape error of the $N_{\rm PC}$-th brightest star.  The
fainter stars do not help very much.  Since $\delta
\hat\xi_{(\ref{error_a})}^{(i,i)}$ contains a sum over the elements of
${\rm Cov}(A)$, it scales similarly.

\section{Multi-exposure Correlation Functions}
\label{multiimage}

In the preceding section, we found two contributions to the systematic
error in the correlation functions which do not scale with the survey
size.  However, notice that both of these, 
$\delta \hat\xi_{(\ref{error_atm})}^{(i,i)}$ and $\delta
\hat\xi_{(\ref{error_a})}^{(i,i)}$ 
only exist for estimates of the
correlation function which use shear estimates from the same exposure
($i=j$).

There is a simple way to eliminate such systematic errors in the
correlation function: for each pair of galaxies used in estimating the
two-point function, use galaxy shapes measured from different
exposures.  In other words, only use pairs with $i \neq j$.  Since the
atmospheric component of $\epsilon_{\rm PSF}$ and that from the errors
in the coefficients are uncorrelated between exposures $i$ and $j$,
there is no systematic bias in the estimates of the correlation
function.  We have thus used the fact that the atmosphere 
has spatial coherence in any given exposure, but gets uncorrelated rapidly
between distinct exposures (as long as they are not taken in 
immediate succession). The same holds for some types of instrumental
systematics which are not correlated between exposures taken on
different nights. The systematic errors eliminated by this 
technique are what we have been calling
$\delta \hat\xi_{(\ref{error_atm})}^{(i,i)}$ and $\delta
\hat\xi_{(\ref{error_a})}^{(i,i)}$.  
So for these two components of the error, $\delta\hat\xi = 0$.

For the three-point (or $n-$point) correlation function, the same
argument holds as long we have at least three (or $n$) exposures. With
the shapes all taken from different exposures, the systematic errors
from the atmosphere and the coefficient estimates are eliminated,
leaving only the errors from the PC measurements and the neglected PCs
to contribute to the systematic error.

By $\delta\hat\xi$, 
we have been referring to a systematic error, or bias; that is,
a change in the expectation value relative to the correct value.
So correlating across multiple exposures results in no 
systematic error from the effects we have numbered \ref{error_atm}
and \ref{error_a}. However, these errors (all four, actually)
also contribute to the statistical error in $\xi$,  
since the variance of $\xi$ due to these two errors does not vanish. 
This contribution to the statistical error has yet to be accurately
estimated, but we expect it to be small compared to the
sample variance plus intrinsic ellipticity errors 
($\epsilon_{\rm intrinsic}$ from Equation~\ref{eqn:psf_error}).

\citet{Wi05} used a set of exposures of a single field imaged with the 
Subaru telescope to estimate the atmospheric contribution. This is a
concern on arcminute scales or smaller, for which the PSF correction
may not be accurate even with PCA interpolation if the PSF pattern is
non-recurrent. \citet{Wi05} finds a contribution to the shear
correlation of order $10^{-5}$ on arcminute scales. This may be compared to the
contribution from intrinsic ellipticities, which is of order
$10^{-3}$, but scales inversely with the total number of galaxy
pairs. The atmospheric contribution scales inversely with the number 
of independent coherent patches,
which depends on the coherence scale of the atmosphere.  Unless this
scale is much larger than an arcminute, the atmospheric contribution
will be comparatively small. The contribution of the other three
errors to the statistical error budget is also likely to be small, but it
needs to be estimated for planned surveys. 

\subsection{Optimal Number of Exposures per Pointing}
\label{exposures}

How many exposures should one take per pointing?  
We have advocated multiple exposures in the discussion above 
to be able to use galaxies in different exposures to measure shear
correlations. While this eliminates certain systematic errors, 
by omitting the $i=j$ terms in the correlation
function estimates, we are losing some information.

Assume each pair of galaxies which are being used for the two-point
correlation function are each observed on $N_{\rm exp}$ exposures and have a
measurement error, $\epsilon_{\rm meas}$, on each exposure equal to
$\sigma \sqrt{N_{\rm exp}}$ (so the measurement error on a stacked
image would be $\sigma$).
The variance of $\xi$ when the 
$i=j$ pairs are neglected is found to be:
\begin{equation}
Var(\xi) = \frac{1}{N_{\rm pair}} \left( 
  \sigma_\gamma^4 + 2 \sigma^2 \sigma_\gamma^2 +
  \frac{N_{\rm exp}\sigma^4}{(N_{\rm exp}-1)^2} 
  \right)
\end{equation}
For well measured galaxies, the last term is negligible (assuming 
$N_{\rm exp} \ge 2$); however, for faint galaxies, it becomes
important.  One generally limits one's measurements to galaxies 
with $\sigma \lesssim \sigma_\gamma \approx 0.25$, since the measurments of
fainter galaxies will often be unstable.  
For $\sigma \approx \sigma_\gamma$, 
we see that the fractional increase in the noise from omitting the $i=j$ pairs
is $N_{\rm exp}/3(N_{\rm exp}-1)^2$, 
which is somewhat significant for only 2 exposures, 
but is small for 5 exposures.

If the shape uncertainties vary significantly between exposures, then our 
approximation that each shear error is $\sigma \sqrt{N_{\rm exp}}$ 
would be incorrect.  A more careful analysis in this case suggests using
enough exposures so that there are at least 2 or 3 with ``good'' 
measurements of the shapes.  For typical variations in the seeing quality,
5 exposures is probably still sufficient.

On the other hand, with large $N_{\rm exp}$, the measurements of the
shapes on each exposure becomes harder relative to the measurement on
a stacked image.  If the signal-to-noise on an individual image drops
to near unity, the measurements may fail to provide any kind of useful
value.  Even signal-to-noise values of 5 or 10 often create problems.
So to avoid having to discard many galaxies which would be measurable
on a stacked image, we definitely want to limit $N_{\rm exp}$ 
to at most 10 or so.

Therefore, we suggest a minimum of 5 and a maximum of about 10 
exposures per pointing for mesurements of shear correlations\footnote{
An absolute minimum of 3 exposures is required to use our 
multi-exposure trick for the three-point correlation functions.}.
For surveys planning to take very many exposures per pointing, 
one would want to stack
subsets of the exposures into 5-10 stacked images and treat these sub-stacks 
as the exposures to which we have been referring.  Sorting the original images
by seeing radius before stacking would probably be the best strategy in 
order to 
not wash out the best-seeing images for at least 2 or 3 of the sub-stacks.

\subsection{Effect on Recurring Principal Components}

The multi-exposure 
technique may also help the first two systematic errors somewhat.
The equations for these errors have terms with $\left\langle a_{ik}
a_{j\ell}\right\rangle$ in them.  With $i \neq j$, the two
coefficients will often be uncorrelated, so this reduces to
$\left\langle a_{ik}\right\rangle\left\langle a_{j\ell}\right\rangle$.
Then, if the coefficients for either component $k$ or $\ell$ have zero
expectation value, then these terms will vanish as well.  Even if the
expectation values are not exactly zero, they may often be much
smaller than the rms, so the $i \neq j$ terms may still be much
smaller than the autocorrelation $i=j$ terms.

Not all coefficients will be uncorrelated between different exposures.
For example, for ground-based telescopes, some components may
correspond to telescope flexure when the telescope points in a
particular direction.  If a given location is always observed at
similar hour angle, the coefficients for these components will be
correlated.

\section{Discussion}
\label{discussion}

This paper has been concerned with the effect of PSF anisotropy
patterns on systematic errors in weak lensing surveys. 
We have suggested the use of galaxy shapes measured in distinct
exposures to estimate shear correlations as a way of eliminating the
systematic error due to non-recurrent PSF patterns. \citet{Ja05}
showed that recurrent PSF patterns can be accurately measured
using a Principal Component Approach. By using these two techniques in
lensing pipelines, systematic errors due to generic PSF patterns can 
be interpolated (and therefore corrected) to high accuracy. 

In planning a large-area cosmic shear survey, we have shown that 
the key factors that enable accurate PSF corrections are: 
sufficiently many well-measured stars in all parts of the sky; 
5-10 exposures per pointing; 
sufficiently few important principal components, which cannot exceed
the number of stars per exposure. In addition, the 
principal components can be
estimated better if dense stellar fields are imaged on regular
intervals, and if there are few changes in the instrument over the course
of the survey (as these can introduce new principal components). 

Another consideration for minimizing the number of important principal
components is to keep the observing conditions as stable as possible.
For each underlying physical cause of PSF variation, one can essentially
do a Taylor expansion of the PSF pattern with respect to that variable.
The PCA will need a separate component for each term in the Taylor
expansion which has a significant amplitude.  Thus, one should try to
keep such variations (eg. focus error, component misalignments, mirror
flexure, etc.) small enough that one or two terms in the expansion are 
sufficient to
adequately describe the effect on the PSF pattern.  One can estimate
what limits are sufficient through spot-diagram ray-tracing programs.

The second goal of this paper was to provide a formalism to estimate residual
systematics due to PSF errors. The ingredients needed to apply 
our formalism are 
an estimate of typical PSF power spectra and of the number of significant 
principal components of PSF patterns. For planned surveys, 
this is best accomplished by generating PSF
patterns in a given exposure by ray tracing through the telescope
optics. Mock surveys can then be generated by modeling the atmosphere 
and the variation of instrumental parameters over the course of the survey. 
The resulting models of PSF patterns can be used with the formalism 
of \S\ref{pca} to find telescope parameters and survey strategy that minimize 
residual systematics. The difficulty in getting reliable estimates of
residual systematics  will be in including all relevant factors which 
may affect the PSF, many of which may be subtle and hard to
anticipate. But the benefit of such an exercise is the ability
to optimize instrument and survey parameters for lensing measurements. 

Further, once data is taken, comparison of the measured principal 
components with the models will help validate the error analysis. 
Our formalism can be applied to survey data to estimate residual 
systematic errors. If systematics turn out to be significant, empirical
estimation allows one to incorporate them in the error budget for 
cosmological parameters. In addition, the following tests provide independent
checks of the estimate of systematic errors from survey data 
(note that at least the latter two
tests can be applied to model PSF patterns for planned surveys as well): 

\begin{itemize}
\item{Stellar ellipticity correlations}

For analysis methods where the corrected stars are not degenerately round, 
the two- and three-point correlation functions of the corrected stars can 
be a measure of how well the interpolation is removing the systematic 
contributions to the correlation.  For a better check, one can perform the 
PSF corrections with only half the stars, and 
look at the resulting correlations of the other half.

\item{Cross-correlation of galaxies with foreground stars}

This will provide a somewhat more direct measure of the contamination from the 
interpolation, since the galaxies use the interpolated PSF.  
Again, one can split the stars in half for a better check.

\item{E/B mode analysis}

The shear field can be decomposed into curl-free (E) and 
divergence-free (B) modes \citep{Sch98, Cr01}.
Most PSF effects have roughly equal power in the E and B-modes, 
while the lensing
signal is (almost) only in the E-mode.  So any residual 
PSF contamination should show up in the B-mode.  However, there are 
some cosmological sources of B-mode power \citep{Cr01, Sch02}, 
so when these become important, this check
will only provide an upper limit to the contamination due to 
the PSF and other systematic effects (e.g. Vale et al 2004).  

\item{Higher order correlations}

Higher order correlation functions, in particular the three-point
function, provide some independent checks on systematic errors. 
The three-point function of the gravitational shear vanishes at
lowest order in the density in the quasilinear regime, so it is non-zero only 
at fourth-order in the density, but it would likely have a third order
contribution from systematic errors. That is, the relative
contribution of systematics could be higher than it is in the
two-point function. Further, there
are multiple three-point functions that contain B-mode contributions,
which would in general behave differently from the two-point
functions. The three-point function also has a shape dependence
that should reveal its gravitational origin and depends somewhat
differently on cosmological parameters than the two-point function 
(Bernardeau et al. 1997; Takada \& Jain 2004). 

\end{itemize}

Finally we note that the methods described here would reduce
systematic errors due to the correction for the size of the PSF 
in addition to those due to the anisotropic PSF described above. 
A round PSF smoothes the images of the galaxies, making them appear
less elliptical.  A multiplicative ``dilution correction''
(part of the shear polarizability in the KSB formalism) 
is therefore needed to obtain the pre-seeing
shape of the galaxies.
PCA interpolation provides better estimates of the size of the PSF, 
which is needed for this correction. And the muti-exposure trick 
would eliminate the contribution of dilution errors due to
non-recurrent patterns in the shear correlation functions. A detailed
study of the resulting improvements is left for further work. 

\acknowledgements We thank David Rusin, Peter Schneider, 
Fritz Stabenau and David Wittman for help and 
comments, and an anonymous referee for suggestions.  
This work is supported in part by NASA
grant NAG5-10924 and by a Keck foundation grant.

\end{document}